\documentclass{PoS}

\title{The nature of gamma ray blazar candidate PMN~J1326$-$5256}

\ShortTitle{The nature of gamma ray blazar candidate PMN J1326$-$5256}

\author{\speaker{Hayley Bignall} and Giuseppe Cim\`o\\
        Joint Institute for VLBI in Europe\\
        E-mail: \email{bignall@jive.nl}, \email{cimo@jive.nl}}

\author{David Jauncey\\
        Australia Telescope National Facility, CSIRO\\
        E-mail: \email{David.Jauncey@csiro.au}}

\author{Cliff Senkbeil, Jim Lovell and Simon Ellingsen\\
        University of Tasmania\\
        E-mail: \email{cliffs@utas.edu.au}, 
                \email{Jim.Lovell@utas.edu.au}, 
                \email{Simon.Ellingsen@utas.edu.au}}

\abstract{A comparison of AGN detected at gamma ray energies by EGRET with
  flat-spectrum radio sources observed in surveys for intraday
  variability reveals that a remarkably high fraction of EGRET
  blazars show significant interstellar scintillation at
  centimetre wavelengths. Scintillating AGN will therefore be targets
  of interest for GLAST, scheduled for launch in early 2008. We
  suggest that the variable, scintillating flat-spectrum radio source
  PMN~J1326$-$5256 is associated with the unidentified EGRET source
  3EG~J1316$-$5244. We describe the properties of PMN~J1326$-$5256 and
  present recent results of monitoring with the ATCA and Ceduna radio
  telescopes.}

\FullConference{From planets to dark energy: the modern radio universe\\
		 October 1-5 2007\\
		 University of Manchester, Manchester, UK}

\begin{document}

\section{Multiwavelength properties of PMN~J1326$-$5256}

There is a strong connection between AGN exhibiting interstellar
scintillation (ISS) and blazars that were detected at gamma ray
energies with EGRET. 
Out of the 19 EGRET blazars observed in the MASIV 5~GHz VLA
Survey \cite{Lovell03,Jauncey07}, 17 showed significant intraday
variability (IDV)
in at least one epoch with rms fractional variations 1--6\%; in
comparison, 56\% of the entire compact, flat-spectrum MASIV sample
showed such IDV. The MASIV survey showed a strong Galactic latitude
dependence of IDV, indicating a predominantly interstellar origin
for IDV at 5~GHz.

The radio source PMN~J1326$-$5256 is a little-studied object with very
few references in the literature, and no optical identification prior
to the present work. It was observed in the ATCA\footnote{The
Australia Telescope Compact Array is part of the Australia Telescope
which is funded by the Commonwealth of Australia for operation as a
National Facility managed by CSIRO.}  calibrator survey and discovered
to be intraday variable (R.J.~Sault 2001, private communication). The
source has no large-scale structure, being completely unresolved with
the ATCA and also the AT-LBA with a maximum resolution of 16~mas at
2.3~GHz.  The most accurate J2000 coordinates available for
PMN~J1326$-$5256 are $13^{\rm h}26^{\rm m}49.23^{\rm s}\pm 0.02^{\rm
s}, -52^{\circ}56'23.7''\pm 0.1''$, determined from ATCA data at
4.8~GHz. This position is coincident with sources in the USNO-A2 and
2MASS catalogues. We obtained an optical spectrum during AAT service
observations on 5 June 2002. The spectrum is featureless across the
observed range of $\sim 5000-9000$\AA, with S/N $\sim 15$ in the
unaveraged continuum. During the AAT observations it was noted that
the object appeared much brighter than on archival UK Schmidt plates,
indicating strong optical variability. Near-infrared colours from
2MASS (J=14.67, H=13.70, K=12.78) match typical BL Lacertae objects.
PMN~J1326$-$5256 is a candidate BL Lac object,
although further optical spectroscopy, particularly in the blue end
of the spectrum, would be useful to confirm this identification.

3EG~J1316$-$5244 is an unidentified EGRET source at an angular
separation of $1.5^{\circ}$ from PMN~J1326$-$5256. This offset is
larger than the quoted error radius of $0.5^{\circ}$, however we note
that 3EG~1316$-$5244 is flagged as having an irregular or not closed
95\% position likelihood contour, indicating possible large
uncertainty in the EGRET source location \cite{Hartman99}. We suggest
a tentative association between 3EG~J1316$-$5244 and PMN~J1326$-$5256
based on multiwavelength properties.

\section{Scintillation monitoring}

An annual cycle in the timescale of interstellar scintillation (ISS) is
expected from the changing velocity of the Earth relative to the
scattering plasma. In principle, observations of ISS at different
times of the year can be used to determine scattering screen
parameters and microarcsecond-scale source structure. 
PMN~J1326$-$5256 was included in an ATCA IDV monitoring programme, in
which it was observed at 4.8 and 8.6~GHz in 14 sessions of $\sim 2$
days or more, over a period of 2.5 years starting in early 2001,
shortly after the discovery of its IDV. This source showed the largest
amplitude IDV of the 21 IDV sources included in the ATCA monitoring
programme, with a maximum modulation index (fractional rms variation)
of 16\% at both 4.8 and 8.6~GHz, although significant IDV was not
observed in every epoch, with the 2-day modulation index being
$\lesssim 1$\% at times.  The MASIV Survey showed that episodic IDV is
a common phenomenon amongst flat-spectrum radio sources \cite{Jauncey07}.

\begin{figure}
\begin{center}
\includegraphics[width=0.5\textwidth]{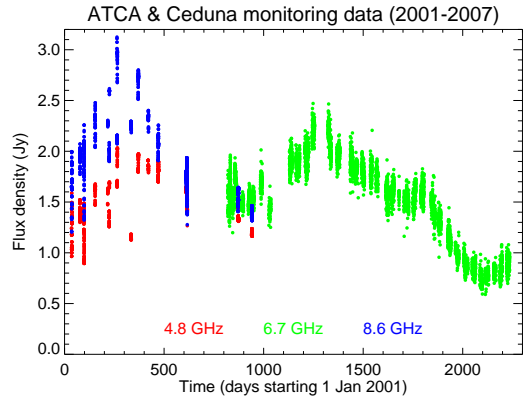}
\caption{Six years of ATCA and Ceduna total flux density monitoring.}
\label{fig1}
\end{center}
\end{figure}

\begin{figure}
\begin{center}
\includegraphics[width=0.55\textwidth]{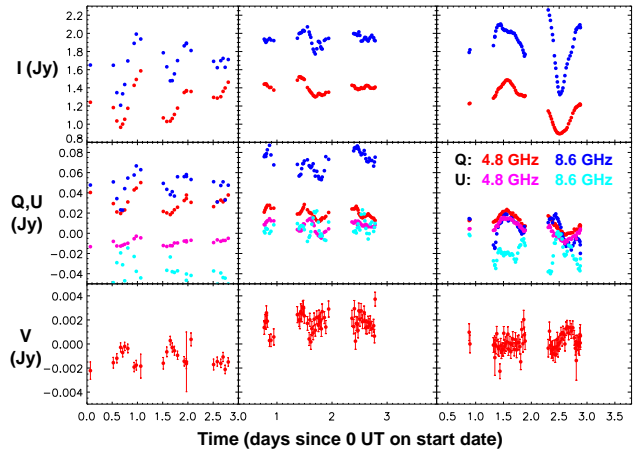}
\caption{First three epochs of ATCA data, starting on 4 February, 17
  March and 6 April, 2001, showing all four Stokes parameters plotted
  with 5-minute or scan averaging.}
\label{fig2}
\end{center}
\end{figure}

PMN~J1326$-$5256 is also a target of the COSMIC project
\cite{McCulloch05}, using the University of Tasmania's 30-m radio
telescope at Ceduna. This programme aims at dedicated monitoring of
several IDV sources at 6.7~GHz. COSMIC observations of
PMN~J1326$-$5256 started in early 2003. Figure~\ref{fig1} shows ATCA
and Ceduna monitoring data up to early 2007. Figure~\ref{fig2} shows
all four Stokes parameters plotted for the first three epochs of ATCA
data. Stokes V (circular polarization) shows a sign flip between the
first two epochs which helps to constrain the origin of the circularly
polarized radiation. The variations at 4.8 and 8.6~GHz are strongly
correlated, indicative of scintillation in the weak scattering regime,
but at Galactic coordinates 
$l=308.3^{\circ}, b=9.6^{\circ}$, PMN J1326$-$5256 would be
expected to undergo strong scattering at these frequencies.
An additional nearby scattering ``screen'' may be
responsible for the rapid IDV observed, since scattering material 
close to the observer has a lower transition frequency and causes more
rapid variations than the same material at a larger distance.

There is some evidence of a slow-down in the ISS timescale around
November from the first year of ATCA data and later Ceduna data, but
no clear repeating annual cycle is seen. The IDV displays large
changes in modulation index (fractional rms variation) at different
epochs. Currently it is not clear whether changes in the IDV behaviour
of PMN~J1326$-$5256 are mostly due to intrinsic source changes or
changes in the interstellar medium along the line-of-sight. Because of
the low Galactic latitude of PMN~J1326$-$5256, the distribution of
scattering plasma along the line-of-sight to this source is likely to
be complex. A superposition of multiple scattering ``screens'' could
lead to multiple scintillation timescales, which are evident in the
Ceduna monitoring data. If the edge of one such scattering region
drifts in or out of the line-of-sight to PMN J1326$-$5256, this could
result in changes in the ISS behaviour. From Figure~\ref{fig1},
however, it is clear that source-intrinsic changes occur on timescales
of months to years. From the ATCA monitoring, the observed decrease in
modulation index with time shows a moderate correlation ($\rho \sim
0.6$) with steepening spectral index, suggesting a possible expansion,
quenching of the ISS and subsequent fading of a compact component.  We
have investigated the radio spectral variability of PMN~J1326$-$5256
using data between 1.4 and 96~GHz from the ATCA calibrator database,
observed irregularly between 2000 and 2007. Taking data from epochs
close in time to estimate instantaneous spectra, there is evidence
that during the period where the source displayed rapid IDV, the
spectrum peaks above 22~GHz, indicating the presence of a strongly
self-absorbed synchrotron component, while at later times during the
``quiescent'' phase observed at Ceduna, the turnover frequency dropped
down below 10~GHz. Recently the spectrum has again started to become
more inverted. Continued monitoring and more detailed analysis and
modelling of the Ceduna light curves are therefore of interest to
determine the origin of the observed changes in the scintillation of
PMN~J1326$-$5256.

\section{Summary}

The radio spectrum, extreme ISS and longer-term intrinsic radio
variability of PMN J1326--5256 imply the presence of a high
brightness temperature, highly beamed jet component.  A first optical
spectrum of the source, together with the photometric properties
indicate that it is a typical low-frequency peaked (classical
``radio-selected'') BL Lac object. We suggest a tentative association
with the unidentified EGRET source 3EG~J1316$-$5244. GLAST will have
the resolution and sensitivity to be able to confirm PMN~J1326$-$5256
as a gamma ray source. Multiwavelength monitoring of sources such as
PMN~J1326$-$5256 in the GLAST era has the potential to make important
contributions to our understanding of the physics of AGN jets and high
energy emission mechanisms.

\end{document}